# Spider Silk: The Mother Nature's Biological Superlens

James N Monks[1], Bing Yan[1] and Zengbo Wang[1]

[1]School of Electronic Engineering, Bangor University, Bangor, UK, LL57 1UT

*Abstract*— **This paper demonstrates a possible new microfiber bio near-field lens that uses minor ampullate spider silk, spun from the Nephila edulis spider, to create a real time image of a surface using near field optical techniques. The microfiber bio-lens is the world's first natural superlens created by exploring biological materials. The resolution of the surface image overcomes the diffraction limit, with the ability to resolve patterns at 100-nm under a standard white light source in reflection mode. This resolution offers further developments in superlens technology and paves the way for new bio-optics.**

## I. Introduction

With enhancing requirements to observe and examine specimens at the nanometer scale, super-resolution optical microscopy methods have been introduced, with such techniques as the microsphere nanoscope [1] assisting in this field. The importance of real-time observations at the nanoscale enables advances within such fields like microbial biology and semiconductor observations. This paper examines the use of minor ampullate spider silk for future developments into bio optics and near-field imaging, by exploring an uncharted optical property of spider silk. This is achieved through computational and experimental observations of the electromagnetic (EM) scattering of white light through the spider silk. The experiment aspect reviews the silk on a sample with known nano dimensions, in order to identify key characteristics of the sample via the imaging performance of the microfiber minor ampullate spider silk. The sample used was a blank commercial Blu-ray DVD disk.

After the invention of the optical microscope in 1609 by Galileo, Ernst Abbe, in 1873, documented the resolution limit of an optical microscope. He established a minimum distance, d, between two compositions to be projected as two objects oppose to a single object. This occurrence was given by;

$$d = \frac{\lambda}{2NA} \quad (1.1)$$

Where λ is the wavelength and NA is the numerical aperture of the objective lens. The minimum distance limitation is associated to the diffraction and loss of evanescent waves within the far field where the sub-wavelength information of the underlying object is situated by carrying the information at a high-spatial-frequency [1]. This information exponentially decays with distance. This gave rise to the Abbe criterion for the fundamental limit of optical microscope resolution.

Research in the late 20th century on super-resolution optical microscopy rapidly grew thanks to the developments of meta-materials, nanophotonics, and plasmonics. This allowed for successful optical nanoscope imaging to be achieved with such techniques involving fluorescence microscopy [2], Pendry-Veselago negative index superlens [3], and Super-oscillatory lens [4]. Although successful at nano imaging, these techniques are at present, restricted to narrow band wavelengths. This makes super-resolution imaging under white-light complicated and difficult to accomplish. However, the advancement in microsphere nanoscope has furthered the field of super-resolution optical microscopy, by resolving sub-diffraction imaging in real time under white light [1]. Microspheres utilise the properties of photonic nanojets to achieve super-resolution foci [5]. An existing issue with microsphere nanoscope was described as not achieving a large field-of-view [6]. The use of cylindrical lenses has the ability to extinguish this limitation as they have the capability to "stretch" and extend the field-of-view [7], and is demonstrated in this research with spider silk being used as a cylindrical lens, as well as outlining a new, additional property which spider silk holds, super-resolution imaging.

Although lenses and optics have previously been influenced by the natural world, with such micro-optics persuaded by the compound eye of insects to create an artificial compact lens system [8], currently there is no literature demonstrating naturally occurring bio superlens. Thus, this paper outlines the world's first superlens which exists by naturally forming in the animal kingdom.

## II. Methodology

With white light focused on the microfiber spider silk, it allowed for photonic nanojet effects to establish the excitation volume as an alternative of the diffraction limited white light foci, leading to enhanced resolution via simulation and experimental investigation. This created real time sub-diffraction images of the features displayed on the commercial blank Blu-ray disk sample. The minor ampullate silk has a refractive index of 1.55 [9], in a surrounding medium with a refractive index contrast relative to the spider silk of less than a ratio of 2:1 [10-12]. The spider silk has a diameter in the range of 6.8µm. The surrounding medium is Isopropyl alcohol (IPA) with a refractive index of 1.377 [12] overlaid with transparent cellulose-based, pressure sensitive tape, which acts as a birefringent retardant [13] as well as allowing for transportation of the bio lens system while providing sound protection for the minor ampullate spider silk.

Figure 1 illustrates the experimental setup schematic of a reflection mode white-light spider silk bio-lens nanoscope. The complementary bio-lens is positioned on the top of the sample surface. A halogen lamp with a peak wavelength of 600 nm is used as the white light illumination source. The bio-lens consists of the minor ampullate spider silk encased within a transparent cellulose-based pressure sensitive tape, with IPA injected between the sample surface and bio-lens. The IPA is present in order to obtain sub-diffraction imaging, by acting as an adhesive force, attracting the mono-fibre bio-lens to the sample of the surface for improved contact. Surface contact is important as the photonic nanojets, which resolve the surface imaging, establish close to the lens surface, thus, surface contact is a necessary for this system to operate.

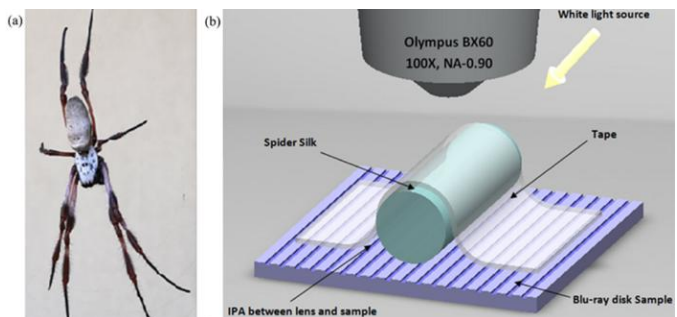

**Fig. 1** (a) Shows an example of the Nephila edulis spider. (b) A schematic demonstrating a complementary microfiber bio-lens integrated with a classical wide field optical microscope under white light. The minor ampullate silk within the bio-lens, collects the near-field object information and projects a surface image that is captured by the conventional lens.

**Minor ampullate spider silk fabrication and preparation for imaging.** The minor ampullate spider silk was spun from a Nephila edulis spider and obtained by the Oxford Silk Group, Department of Zoology, University of Oxford. The fabrication and preparation of the silk for imaging purposes was conducted under a controlled environment within a laboratory. A strand of silk was carefully handled and placed directly onto a clean and uncontaminated glass, where the silk was then overlaid with a surrounding medium of transparent cellulose-based, pressure sensitive tape, creating the mono-bio-flexible lens. This surrounding medium allows for careful directional manipulation along the x and y axis of the imaging sample.

**Computational method.** The geometrical optical analysis and field distributions in figure 2 were calculated through Mie theory and finite integral techniques using CST microwave studio.

III. SUPER-RESOLUTION IMAGING

**Super resolution mechanics.** Applying high-resolution finite integral technique (FIT) for computational solutions of Maxwell's equations, presented evidence that a plane wave of white light illumination directed onto a circular dielectric cylinder can create a narrow, high-intensity, sub-diffraction waist beam that transmits into the surrounding medium [9]. The circular dielectric cylinder has been adapted for the minor ampullate spider silk in order to obtain the mechanics surrounding the super-resolution ability.

The importance of 'photonic nanojet' (PNJ) for optical imaging has been will established; with such demonstrations performed using microsphere nanoscopy, which is able to overcome the diffraction limit to resolve patterns at 50 nm resolution under white light with the ability to magnify up to 8x [1].

The minor ampullate spider silk exploits PNJ technique to accomplish super-resolution focusing. Figure 2b exposes the fundamental traits of PNJ in the minor ampullate. The electric field is significantly improved in the near-field zone under the silk, with a near exponential decay along the light propagation z-direction, which is shown in figure 2a along with the peak intensity of the PNJ.

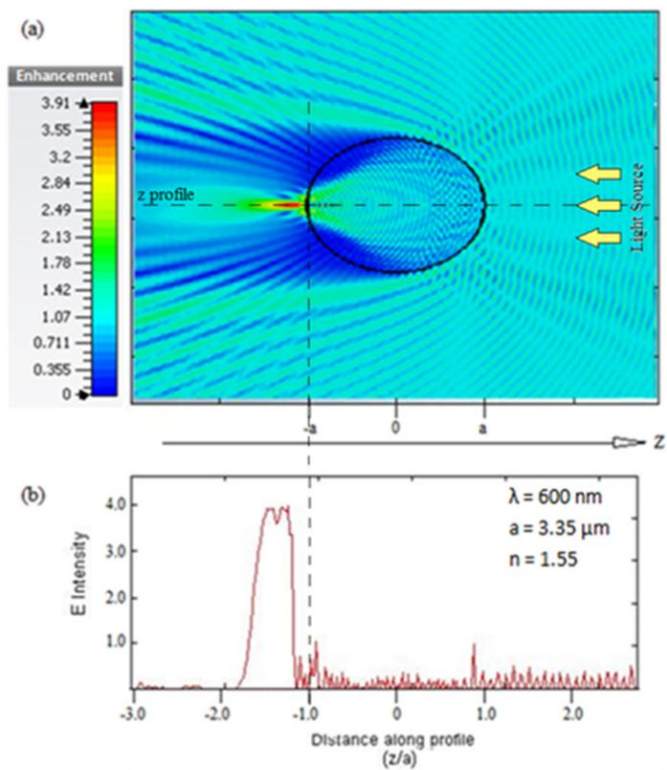

**Fig. 2** Photonic nanojets spatial intensity distribution produced by the minor ampullate spider silk inside and outside the 6.7 µm diameter fibre, illuminated by a plane wave at λ = 600nm. (a) Shows the cross sectional area of high intensity, sub-diffraction waist beam under the spider silk fibre within a surrounding medium of air to demonstrate the photonic nanojet establishment. (b) Shows the intensity along the z-axis, z = -1.0 is the position under the spider silk fibre.

**Experimental imaging performance.** Figure 3b presents the typical optical performance of the minor ampullate spider silk. The microfiber collects the underlying object information, magnifies and projects the image to the far field. The image was composed by a conventional 100x objective lens, numerical aperture NA = 0.90. The image was captured using a 6.8 µm diameter thread of minor ampullate spider silk. The figure resolves a Blu-ray DVD disk, 200 nm wide channels separated 100 nm apart, imaged in reflection mode using a halogen light illumination and clearly resolve the sub-diffraction pattern. This provides evidence of the super-resolution capability of spider silk to overcome the optical diffraction limit.

The resolved Blu-ray pattern image in figure 3 is situated to the right of the centre plane of the silk. This is due to the incident light being delivered from the opposed direction forcing the light radiation to scatter and reflect the surface image in the right hand side of the spider silk.

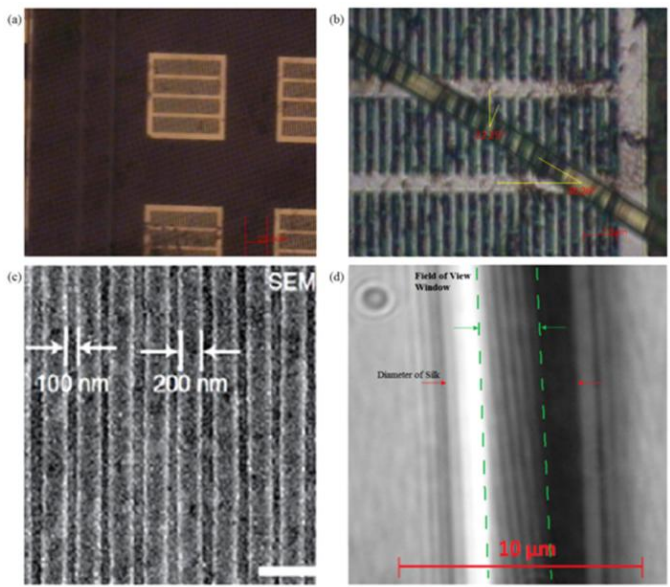

**Fig. 3** A typical imaging example of mono-bio-flexible lens nanoscope in reflection mode imaging a surface of a integrated circuit and commercial Blu-ray DVD disk. The 100-µm thick transparent protection layer of the Blu-ray disk has been removed previous to using the minor ampullate spider silk (diameter of 6.8 µm). Micro dimensional integrated surface pattern (a; Optical imaging) is magnified by the spider silk (b; Optical Nanoscope image). The sub-diffraction 100 nm channels (c; SEM image) are resolved by the spider silk superlens (d; Optical Nanoscope image) and corresponds to a magnification fact of 2.1X.

Further evidence for the super-resolution capacity of spider silk is presented in figure S1. A Barium Titanate ($BaTiO_3$) microsphere, with a similar diameter to the spider silk, was located next the silk strand and encased within a surrounding medium of Polydimethylsiloxane (PDMS). Microspheres have the clear ability to resolve sub-diffraction patterns [1,5], with figure 4 plainly distinguishing the 200 nm wide channels and the 100 nm separation lines. Similar to figure 3d, the spider silk in figure S1 unmistakably resolves the pattern while under reflection mode. However, the spider silk projects the pattern at an angled image. The angled image is expected and can be described by the laws of refraction and convex lens systems. The focal lens plane of the spider silk differs from that of the microspheres, hence why the microsphere resolution is slightly out of focus.

The true direction of the Blu-ray pattern is resolved by the microsphere, which could be used as a reference point. The incident wave light passes through the silk creating a PNJ, which stored the surface information and is reflected back into the spider silk and finally back to air to be picked up by the optical microscope. The surface information light waves reflected back through the silk are refracted to a greater degree when the information reaches the air, forcing the resolved image to distort. This angle adjusts depending on the direction of the silk strand in relation to the direction of the surface sample pattern. As the centre plane angle of the spider silk increases from the directional angle of the sample pattern, the resolved image angle is also increased resulting in greater distortion.

## IV. DISCUSSION

To summarise, this paper verified that the minor ampullate spider silk, spun from the Nephila edulis spider, has the properties to perform as an optical superlens by utilising photonic nanojets to carry the surface information at a high-spatial-frequency, which can resolve 100-nm objects and patterns under a white light source illumination. The mono-bio-flexible lens nanoscope, with future developments, could become an easy integration enhancement device to optical microscopes for reflection mode observations, forming a super-resolution optical nanoscope with real time imaging. The silk acts as a cylindrical lens and has the ability to overcome the field-of-view limitation that the microspheres present. The spider silk nanoscope is robust and economical, providing a minimum manufacturing platform for future commercial endeavours. This type of lens is the first biological superlens system that has successfully overcome the diffraction limit.

The current constraints of the lens system proposed comprises of the intricate contact of the minor ampullate spider silk with the surface of the imaging sample. Additionally, the lens projects distorted images of the surface sample, requiring the need for either knowledge of the true surface pattern direction, or the need of a reference point built into the lens system.

**Future work.** The review of contact relationship between the minor ampullate spider silk and the surface of the imaging sample could be explored further in order to achieve greater contact, thus, potentially improving the super resolution imaging performance. Prospected developments for consideration incorporate a multi-bio-flexible lens, where the minor ampullate spider silk is arranged in a two dimensional array with a section of microspheres for reference, coated in a surrounding medium of transparent cellulose-based, pressure sensitive tape, or an improved surrounding material such as PDMS. Additionally, the spider silk could potentially be doped with $BaTiO_3$ in order to achieve a cylindrical lens that could closely match that of the $BaTiO_3$ microspheres resulting in the extension of the field-of-view.

## ACKNOWLEDGMENT

This investigation would not be possible without the generosity and backing from Professor Fritz Vollrath and his research team from the Oxford Silk Group, Department of Zoology, University of Oxford.

## SUPPLEMENTARY MATERIAL

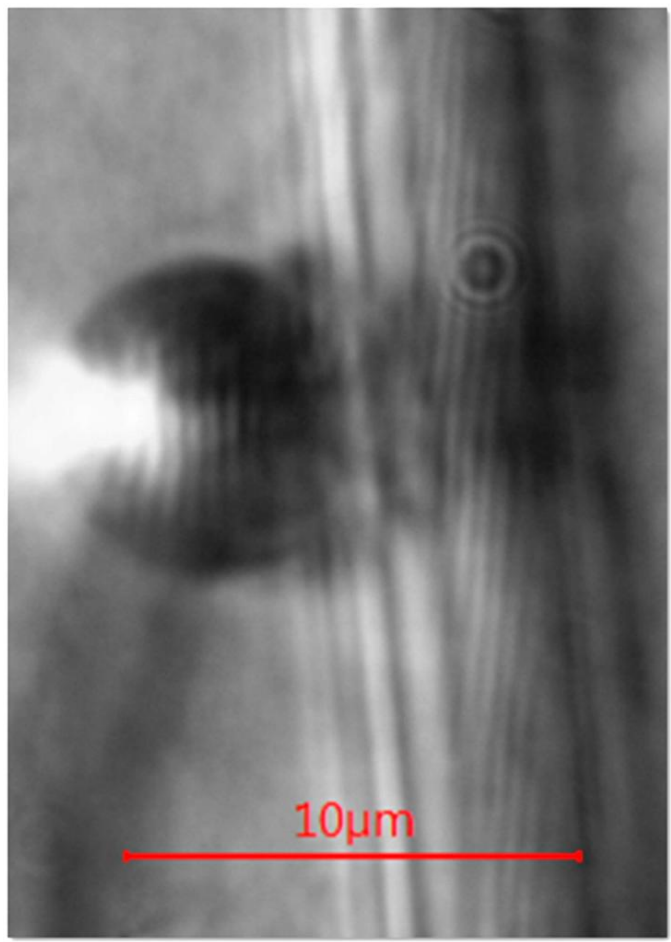

**Fig. S1** Microsphere positioned beside the minor ampullate spider silk in reflection mode, images a commercial Blu-ray DVD disk. The image resolved within the spider silk is rotated by approximately four degrees compared to the real direct of the sub-diffraction 100 nm channels shown by the microsphere.